\documentclass{aa}  
\usepackage{amsmath}
\usepackage{graphicx}
\usepackage{txfonts}
\usepackage{natbib}
\usepackage{color}
\usepackage{ulem}

\graphicspath{{figure/}}

\begin{document}
   \title{Pulsations of rapidly rotating stars\\ II. Realistic modelling for intermediate-mass stars}
 \titlerunning{Pulsations of rapidly rotating stars}

   \author{R-M. Ouazzani
          \inst{1}
          \and
          I. W. Roxburgh\inst{2}
          \and
          M-A. Dupret\inst{3}
          }

   \institute{Stellar Astrophysics Centre, Department of Physics and Astronomy, Aarhus University, Ny Munkegade 120, DK-8000 Aarhus C, Denmark. 
              \email{rhita-maria.ouazzani@phys.au.dk}
         \and
             Astronomy Unit, Queen Mary University of London, Mile End Road, London, E1 4NS, UK
         \and
         Institut d'Astrophysique et de G\'eophysique de l'Universit\'e de Li\`ege, All\'ee du 6 Ao\^ut 17, 4000 Li\`ege, Belgium }

   \date{Received January 21st / Accepted May 4th}

 
  \abstract
   {Very high precision seismic space missions such as CoRoT and Kepler provide the means  for testing the modelling of transport processes in stellar interiors. For some stars, such as $\delta$ Scuti $\gamma$ Doradus and Be stars, for instance, the observed pulsation spectra are modified by rotation to such an extent that it prevents any fruitful interpretation.}
   {Our aim is to characterise acoustic pulsation spectra of realistic stellar models in order to be able to interpret asteroseismic data from such stars.}
   {The 2-dimensional oscillation code ACOR, which treats rotation in a non-perturbative manner, is used to study pulsation spectra of highly distorted evolved models of stars. 2D models of stars are obtained by a self-consistent method which distorts spherically averaged stellar models a posteriori, at any stage of evolution, and for any type of rotation law. }
   {Four types of modes are calculated in a very dense frequency spectrum, among which are island modes. The regularity of the island modes spectrum is confirmed and yields a new set of quantum numbers, with which an \'echelle diagram can be built. Mixed gravito-acoustic modes are calculated in rapidly rotating models for the first time.}
   {}

\maketitle

\section{Introduction}
Rapid stellar rotation introduces a number of phenomena which considerably complicate the modelling of stars and the study of their pulsation modes. On the one hand, centrifugal acceleration reduces local gravity, mimicking a lower mass. On the other hand, rotation induces meridional circulation \citep{Eddington1925} and shear and baroclinic instabilities \citep{Mathis2004}, which contribute to the mixing of chemical elements. Although crucial for modelling stellar structure and evolution, the problem of transport of angular momentum inside stars has not yet been fully understood. 

As shown with great success for evolved low mass stars, asteroseismology is a very efficient tool to constrain the internal structure of stars. Unfortunately, for intermediate mass stars, because they do not undergo magnetic braking, rapid rotation comes into play and makes the interpretation of their seismic spectra much more delicate.   
Fast rotation is known to have a strong impact on stellar pulsations: the centrifugal force distorts the pulsations cavity and the Coriolis force modifies the dynamic of modes. So far the impact of fast rotation on observational frequency spectra is poorly understood, and some questions still need to be answered. What are the modes corresponding to each observed frequency? Which cavity inside the star do they probe? How can the physics inside this cavity be extracted? All these questions call for realistic modelling.

For slow rotators, the effect of rotation are  taken into account as a small perturbation of pulsations (see for instance \citealt{Ledoux1951} for first-order effects, the second-order effects have first been explored by \citealt{Vorontsov1981,Vorontsov1983}, see also \citealt{Dziembowski1992}, and we refer to \citealt{Soufi1998} for third-order effects). But these methods, although elegant and simple to use, have their limitations \citep{Reese2006,Suarez2010};  they  fail to adequately model the effects of rapid rotation in stars that show very high surface velocities, such as $\delta$ Scuti and Be stars, nor  
 the impact of the Coriolis force on gravity modes in stars whose surface rotates slowly, but in which the pulsation periods are of the same order as their rotation period ($P_{rot}\sim P_{osc}$), such as SPB  or $\gamma$ Doradus stars

This is important for many stars in the CoRoT and Kepler fields of observation. Hence, if one aims at correctly extracting the internal structure from seismic observations, they need to correctly describe the effects of rotation on pulsations. In that context, the aim of this article is to investigate synthetic seismic spectra of such stars for any type of rotation profile $\Omega(r,\theta)$, differential both in radius $r$ and co-latitude $\theta$, and to propose new seismic diagnostics in order to interpret their seismic observations.  



In a previous paper \cite{Roxburgh2006} presented self-consistent 2-dimensional models of main-sequence stars where the angular velocity is differential both in depth and in latitude, and for different stages of evolution. The adiabatic oscillations of stars are only governed by the acoustic structure of the star, that is the hydrostatic structure: density, pressure, gravity and adiabatic exponent. The pulsations do not depend on whether the star is in thermal equilibrium. The self-consistent method allows one to determine such an acoustic structure. 

In the first paper \citep[][hereafter Paper I]{Ouazzani2012b}, we presented the ACOR pulsation code --Adiabatic Code of Oscillation including Rotation--, and its results for a 2-dimensional polytropic model. Here, we report on pulsation modelling of an evolved distorted model of an intermediate mass star, typical of a $\delta$ Scuti star. Section \ref{S_Dist} introduces the method applied to obtain distorted models of stars following \cite{Roxburgh2006}. Section \ref{S_Puls} is dedicated to the pulsation modelling with an emphasis on how the ACOR code was improved in order to compute the pulsations of realistic distorted stellar models. In Section \ref{S_Ex} we present an application to a typical model of intermediate mass star. Finally, Sect. \ref{S_Reg} reports on regularities revealed in the theoretical pulsation spectrum of the model, and conclusions are given in Sect. \ref{S_CCL}.


\section{Self-consistently distorted model}
\label{S_Dist}
Stellar evolution in two dimensions has yet to be modelled. Progress is being made in this field for instance by the ESTER project (for \textit{Evolution STEllaire en Rotation,} \citealt{Rieutord2007,EspinosaLara2013}). The ultimate goal of the ESTER project is to solve the hydrodynamics equations in two dimensions along with the chemical evolution. Currently ESTER can only model the steady state with regards to evolution, the central chemical composition being specified ad hoc.

An alternative to this approach is to calculate spherically averaged stellar evolution models, then distort the models \textit{a posteriori} using a self-consistent method. The first self-consistent model was developed for rapidly rotating polytropes and white dwarfs by \cite{James1964}. See also \cite{Hurley1964}, closely followed by a two-zone model by \cite{Monaghan1965} and two-zone models for rapidly rotating main-sequence stars by \cite{Roxburgh1965}. In parallel, \cite{Ostriker1968} had developed and used such an approach to compute uniformly and differentially rotating polytropic stellar models. The method has then been improved by including more realistic input physics \citep{Jackson1970}, but only applied to massive stars, as convergence problems arose when dealing with low mass stars, due to the high concentration of the central mass \citep{Clement1978}. \cite{Jackson2004,MacGregor2007} solved this problem and successfully applied the method to intermediate mass stars. In their models, the only remaining limitation is the chemical homogeneity and conservative laws of rotation.

We present here such an approach based on a method established by \cite{Roxburgh2006} which allows the computation of non-homogeneous, fully differentially rotating stellar models. It consists of building the two-dimensional acoustic structure of a rotating star in hydrostatic equilibrium starting with spherical profiles of the structural quantities using an iterative scheme. 
 
\subsection{Acoustic structure}

We recall that the adiabatic pulsations of stars are solely governed by their acoustic (i.e.\ hydrostatic) structure, that is the internal density $\rho$, pressure $p$ and gravity $\nabla \Phi$, together with the adiabatic exponent $\Gamma_1$. 
As shown in \cite{Roxburgh2006}, for rotating stars the 2-dimensional hydrostatic structure $\rho(r,\theta)$, $P(r,\theta)$ and gravitational potential $\Phi(r,\theta)$ are determined solely by the angular velocity profile  $\Omega(r,\theta)$, the density $\rho_m(r) = \rho(r,\theta_m)$ and the surface pressure $P_s$ along any one angle $\theta_m$. The adiabatic exponent $\Gamma_1(r,\theta)$ is determined from $\rho(r,\theta)$, $P(r,\theta)$ and the 2-dimensional hydrogen profile $X(r,\theta)$.  If we assume a particular model of chemical elements transport (which is the case in this work: $X$ constant on surfaces of constant entropy) then it is sufficient to have $X_m(r)$, along $\theta_m$. 
The method then consists of generating acoustic models of rotating stars with any $\Omega(r,\theta)$ by considering that the radial density and hydrogen abundance along the angle $\theta_m$ take the values from a spherically averaged model, yielded by 1-dimensional evolution modelling. The reference angle $\theta_m$ is taken here as the root of the second order Legendre polynomial: $\theta_m = \cos^{-1}(1/\sqrt 3)$.     

\label{Ss_acoustic}
\noindent The scheme is the following:\\
\indent 1. Take as an initial guess  $\rho(r,\theta) = \rho_m(r)$\\
\indent 2. Solve the Poisson equation: 
\begin{equation}
\nabla^2 \Phi = 4 \pi G \rho 
\end{equation}
 for $\Phi(r,\theta)$ given $\rho(r,\theta)$, together with the proper boundary conditions.  This is done by expanding $\Phi(r,\theta)$ and $\rho(r,\theta)$ into series of Legendre polynomials. The density decomposition is found by imposing that  $\rho(r,\theta) = \rho(r_i,\theta_j)$ at each grid point ($r_i,\theta_j$). Given that, the coefficients of the gravitational potential expansion are determined  from an integral representation of the solution of the Poisson equation and the upper boundary condition projected onto the Legendre polynomials \citep[for a detailed solution, see][]{Roxburgh2004}. \\
\indent 3. Solve the curl of the hydrostatic equilibrium:
\begin{equation}
\overrightarrow{\rm curl} \left(\rho \left[ \overrightarrow{\nabla} \Phi - \Omega^2 \overrightarrow{\varpi} \right] \right) = 0
\label{curlHE}
\end{equation}
 for  $\rho(r,\theta)$ given $\Phi(r,\theta)$ from step 2. $\overrightarrow{\varpi} = (r \sin \theta, r \cos \theta, 0)$ is the vectorial distance from the rotation axis. This is done in two steps: first, we seek for the characteristics curves $r(\theta)$  along which the partial differential equation is an ordinary differential equation (ODE), and then solve Eq. (\ref{curlHE}) as an ODE along these curves \citep[for a detailed solution, see][]{Roxburgh2006}. \\
\indent 4. Then this density profile is injected at step 2 and steps 2 and 3 are iterated until convergence of the equatorial and the polar radii.\\
\indent 5. The characteristic curves turn out to be the isobars. Once the characteristic curves are determined, all we need is the value of $P$ along a radius at any angle $\theta$, $P(r,\theta)$ is then constant along the characteristic curves.\\
\indent 6. To determine the 2-dimensional profile of the adiabatic exponent given  $P(r,\theta)$ and $\rho(r,\theta)$, we assume that $X(r,\theta)$ is constant on isobars (justification for this assumption follows in Sect. \ref{Ss_discussion_dist}). We then solve the equation of state for $\Gamma_1 = \Gamma_1(P,\rho,X_m)$, $X_m$ being the value of the hydrogen mass fraction at the reference angle $\theta_m$. This gives $\Gamma_1$ along the isobars.
For further details we refer to \cite{Roxburgh2006}.

\subsection{Discussion}
\label{Ss_discussion_dist}

For the algorithm to converge, several approximations must be assumed, the first of which concerns the structure of surface layers. In stars that are differentially rotating, the fluid is baroclinic. This means that surfaces of constant pressure do not correspond to surfaces of constant density. Hence the solution of the algorithm presented in the previous section may have $\rho = 0$ on surfaces of $P > 0$. For the solution to be well behaved at the surface, we need ${\bf curl} (\Omega^2\,\overrightarrow{\varpi}) = 0$ in the surface layers. This is ensured by taking $\Omega$ constant in the outer layers of the star.

Moreover, in this region of low density and pressure, the meridional circulation driven by the rotation can become very large, and induce small scale turbulence. We assume here that the time scale of the circulation is very small in these layers, so that transport of angular momentum is rapid, and the system reaches a steady state ($\Omega$ constant near the surface).

Finally, in order to retrieve the adiabatic exponent $\Gamma_1$, we have assumed that the hydrogen mass fraction $X$ is constant on isobars. However, in the fully ionised interior of most stars where, due to nuclear reactions, $X$ differs from its initial value $X_0$, the values of $\Gamma_1$ are only weakly dependent on $X$. On the other hand in the outer layers, where $\Gamma_1$ can vary considerably, in absence of diffusion $X$ is unchanged from its initial value $X_0$. Therefore $X$ constant on isobars seems to be a reasonable assumption here.


\section{Pulsations modelling}
\label{S_Puls}

As mentioned above, for the computation of pulsations we only require the hydrostatic structure of the star and not the thermal structure. We compute the oscillation modes as the adiabatic response of the structure to small perturbations -- i.e. of the density, pressure, gravitational potential and velocity field -- using the Eulerian formalism.

The code ACOR has been developed for this purpose and is presented in Paper I. Here we report on new improvements which were necessary to implement in order to compute the pulsations of non-barotropic models of stars, which present steeper gradients around the poles than near the equator.

\begin{figure}[t!]
\centering  	    {\includegraphics[scale=0.38, angle=-90]{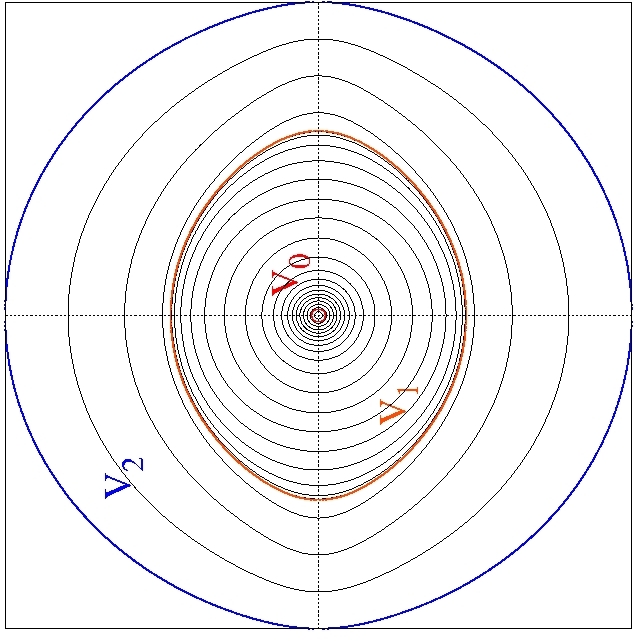}}
            \caption{\label{Fig_CoordSyst}Coordinate system used in ACOR. V$_0$ corresponds to a region inside the convective core,
V$_1$ extends from a boundary inside the convective core to the stellar surface, and V$_2$ encompasses the star.}
\end{figure}

\subsection{A new coordinate system}
The method reported in Sect. \ref{S_Dist} yields a 2-dimensional stellar model described on the characteristics curves (isobars). In order to avoid potential uncertainties caused by interpolation, the idea is to keep the structural quantities expressed on the isobars as much as possible when computing the pulsations of such models. 
Nevertheless, to simplify boundary conditions --at the centre, where the eigenfunctions should behave spherically, and at infinity, where the gravitational potential should vanish-- a new hybrid coordinate system is proposed. The new radial coordinate $\zeta$ is expressed as a bijective, monotonic and continuously differentiable function of the radius $r$. From spherical coordinates at the centre, the system progressively matches isobars, and from the distorted stellar surface becomes spherical again outside the star. Therefore, we adopted a multidomain approach, which consists in dividing the physical space into 3 domains whose boundaries correspond to the model’s discontinuities: one  central domain $V_0$ whose upper boundary is included in the convective core, one domain $V_1$ which encloses the stellar surface, and one external domain $V_2$, bounded by the stellar surface and an external sphere.\\

\noindent \underline{Domain $V_0$} from $\zeta = 0$ to $\zeta = \rm x_{c,eq}$\\
Here $\rm x_{c,eq}$ is the ratio of the radius at the edge of the convective core over the total radius of the star, both at equator (see Fig. \ref{Fig_CoordSyst} the inner domain, encircled by the red boundary). In Paper I, where polytropic structural models were treated, the bijective transformation continuously goes from spherical at the centre to matching the stellar surface without intermediately adopting the isobars.
In the case of a non-barotropic model of star, isobars are closer to each other at the poles than at the equator. To ensure continuity of $\partial_{\zeta} r$ over the boundary $\zeta = \rm x_{c,eq}$, $\partial_{\zeta} r$ should depend on the latitude when $\zeta$ goes to $\rm x_{c,eq}$. 
Hence, the following transformation has been developed:
\begin{align}
\label{Eq_zeta_centre}
\forall \, \zeta_{\rm c} = \zeta / {\rm x_{c,eq}} \, \in \,& \left[0;1\right] \, , \forall \, \theta \, \in \, \left[0;\pi\right] \, , \\ 
\rm x(\zeta,\theta) \, = \, {\rm x_{c,eq}}& \, \left[ \zeta_{\rm c} \, + \, \frac{5 \zeta_{\rm c}^3-3\zeta_{\rm c}^5}{2} \, \left( \frac{\rm x_c(\theta)}{\rm x_{c,eq}} - 1 \right) \right. \nonumber \\
&+ \, \left. \frac{\zeta_{\rm c}^5-\zeta_{\rm c}^3}{2} \,  \left( \frac{1}{\rm x_{c,eq}} \, \frac{\partial x_c}{\partial \zeta}(\theta) - 1 \right) \right]\, , \nonumber 
\end{align}
where x is the fractional radius ($r/R$), and ${\partial x_c}/{\partial \zeta}$ is the derivative of x with respect to $\zeta$ at $\zeta = \rm x_{c,eq}$.
The first term in brackets enforces a spherical behavior for $x(\zeta,\theta)$ near the center, the second term allows to match the isobar $x_c(\theta)$ located at $\zeta_c = 1$ and the last term insures that the derivative of $x(\zeta,\theta)$ at $\zeta_c = 1$ matches $\partial_{\zeta} x_c(\theta)$. We chose the simplest functions which fulfilled the constrains. Odd polynomials of $\zeta_c$ have been chosen because they go faster toward zero at the centre.
In practice, $\rm x_{c,eq}$ can take any value smaller than the actual fractional radius of the $\mu$ gradient location, both for low mass and massive stars. \\

\noindent \underline{Domain $V_1$} from $\zeta = \rm x_{c,eq}$ to $\zeta = 1$\\
In this domain, constant $\zeta$ correspond to the isobars, with $\zeta = x(\theta_m)$. $x(\zeta,\theta)$ is determined by the characteristics equation.\\

\noindent \underline{Domain $V_2$} from $\zeta = 1$ to $\zeta = 2$\\
In $V_2$ the coordinate system goes gradually from the distorted stellar surface to a sphere. A new transformation is proposed, which ensure the continuity of $r$ and $\partial_{\zeta} r$ at the surface of the star $\zeta = 1$. The same mathematical form as for Eq.(\ref{Eq_zeta_centre}) applies here, except that Eq.(\ref{Eq_zeta_ext}) is not taken in zero, therefore, there is no need for odd polynomials. The simplest form which fulfills the constrains is:

\begin{align}
\label{Eq_zeta_ext}
\forall \, \zeta \, \in \, \left[1;2\right] \, , &\forall \, \theta \, \in \, \left[0;\pi\right] \, , \\ 
\rm x(\zeta,\theta) \, = \, \zeta \, &+ \, (2 \zeta^3-9 \zeta^2+12 \zeta -4) \, \left( {\rm x_s(\theta)} - 1 \right) \nonumber \\
&+ \, (\zeta^3-5 \zeta^2+8 \zeta-4)  \,  \left( \frac{\rm d x_s}{\rm d \zeta}(\theta) - 1 \right) \nonumber 
\end{align}

In the two domains V$_0$ and V$_2$ where the coordinate system does not rely on isobars, the structural quantities are interpolated on the new grid ($\zeta, \theta$) by the help of cubic polynomial interpolation over two grid points.

\subsection{The pulsations computations}

It consists of solving the hydrodynamics equations perturbed by Eulerian fluctuations, by direct integration of the 2-dimensional problem. The numerical method is based on a spectral multi-domain method which expands the angular dependence of eigenfunctions into spherical harmonics series, and whose radial treatment is particularly well adapted to the behaviour of equilibrium quantities in evolved models (at the interface of convective and radiative regions, and at the stellar surface). The radial differentiation is made by means of a sophisticated finite difference method which is accurate up to the fifth order in terms of the radial resolution \citep[developed by][]{Scuflaire2008}.

This code has been validated by comparison with the results of \cite{Reese2006} for polytropic models. The agreement between the two codes is found excellent. All details concerning the ACOR code are given in Paper I.


\section{An example, $M = 2 M_{\odot}$, $X_c = 0.35$, $\Omega = \Omega(r)$}
\label{S_Ex}

\subsection{The stellar model}
\begin{figure}[t!]
\centering  	    {\includegraphics[scale=0.3, angle=-90]{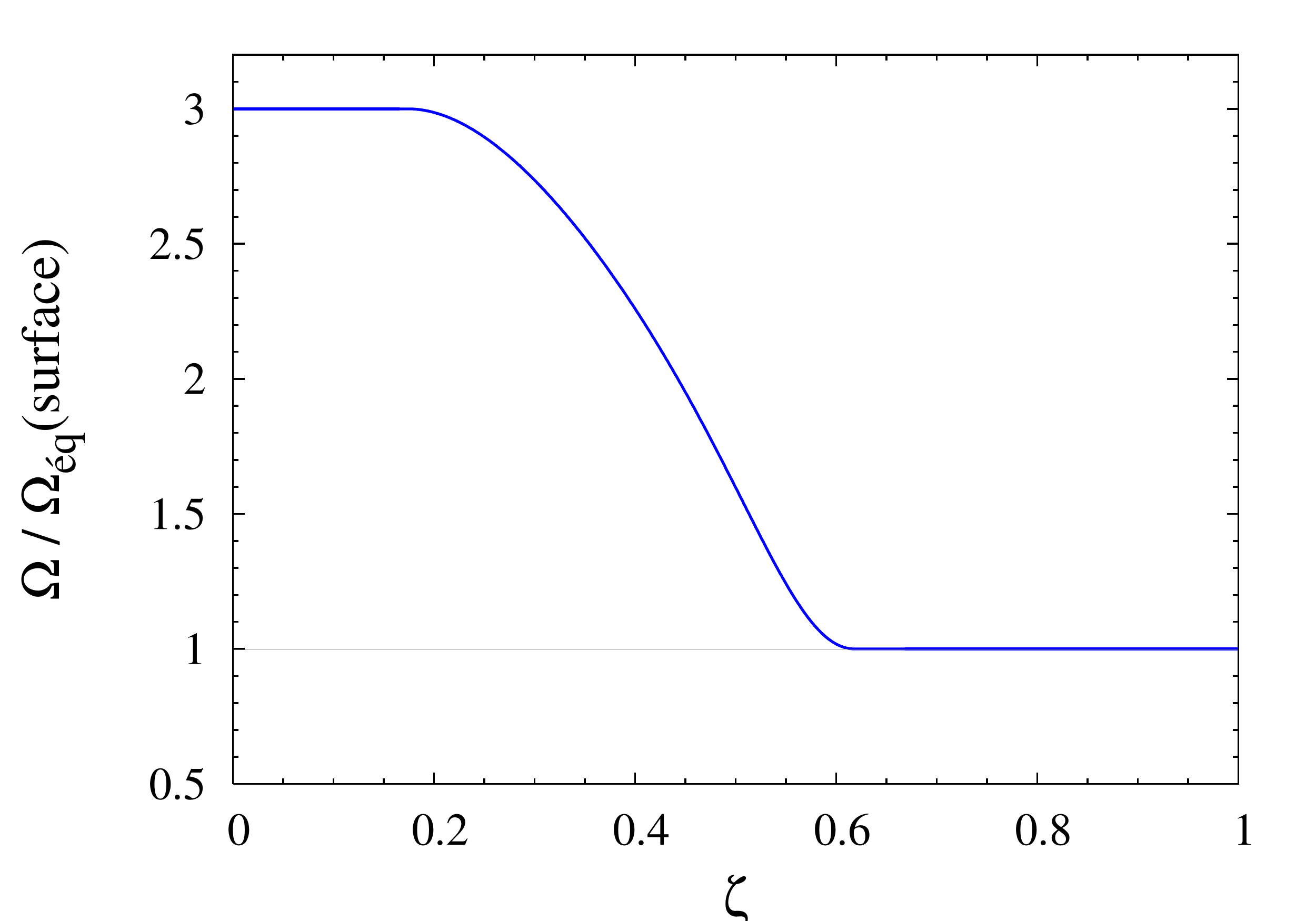}}
            \caption{\label{Fig_rotprof}Radial profile of the rotation angular velocity scaled by the equatorial angular velocity at the surface. $\Omega$ is constant on isobars but differential in the normal direction. Therefore, the model is not barotropic.}
\end{figure}

As an illustrative example we take $\Omega = \Omega_s \omega(x)$ as a function only of radius $x = r/R_0$, where $\omega(x)$, shown in Fig. \ref{Fig_rotprof}, decreases from a value of 3 in a central core to 1 in the outer envelope, and has continuous derivatives. As mentioned in Sect. \ref{Ss_discussion_dist}, with $\Omega$ constant in the outer layers the solution is well behaved at the surface. For the spherically averaged model we took a star of 2 M$_{\odot}$ with an initial composition X = 0.72, Z = 0.02, evolved to the stage where X$_c$ = 0.35. The average centrifugal force $2\Omega^2r/3$ was added to the hydrostatic equation, the remaining equations being those of spherical star. The dimensionless angular velocity $\omega(x)$ was taken as fixed throughout the evolution and $\Omega_s$ determined by requiring conservation of angular momentum for the star as a whole. The initial model had radius R $\simeq$ 2.5 R$_{\odot}$ and $\Omega_s = 4.3 \times 10^{-5} \rm rad.s^{-1}$ corresponding to an equatorial velocity of $\simeq 80\,\rm km.s^{-1}$. The star was evolved to a central hydrogen abundance X$_c$ = 0.35, at which stage $\Omega_s = 1.296 \times 10^{-4} \rm rad.s^{-1}$.

The models were produced using the STAROX evolution code \citep[cf.][]{Roxburgh2008} with the OPAL2001 equation of state \citep{Rogers2002}, OPAL GN93 \cite{Iglesias1996} and \cite{Alexander1994} opacities, NACRE \citep{Angulo1999} nuclear reaction rates, and a radial mesh in mass with N = 2000 points. The spherically averaged model was taken to represent the 2-dimensional model along the angle $\theta_m = \cos^{-1}(1/\sqrt{3})$, and the procedure described in Sect. \ref{Ss_acoustic} followed to produce the 2-d model. The angular mesh had N$_j$ = 240 points and the mesh in domain V$_2$ for solving Poisson’s equation had N$_k$ = 500 points. The flatness of the model $1 - \rm R_{pol}/R_{eq} \simeq$  0.25, the equatorial rotational velocity $V_{eq} = 282.3 \,\rm km.s^{-1}$, and the ratio of centrifugal force to gravity at the surface equator $\Omega_s^2 \rm R_{eq}^3 / G M  = 0.65 $.

\subsection{The pulsation modes}

Given the rotation velocity ($\Omega = 0.81 \Omega_k$ at the surface, where $\Omega_k = \sqrt{GM/R_{eq}^3}$) a convergence study found that 40 spherical harmonics were necessary in the spectral expansion of eigenmodes. In the present article we confine our analysis to axisymmetric modes, although there is no restriction to such modes in the ACOR code. Odd and even modes with $m=0$ are presented here. The parity of a mode $p$ is given by $p = (\ell+|m|)$ mod 2. Therefore modes with $p=0$ are called even modes, and modes with $p=1$ odd modes. For axisymmetric modes, odd modes (resp. even) are expanded on spherical harmonics series with only odd (resp. even) values of angular degree $\ell$.  

\begin{figure}[t!]
\centering 
  	    {\includegraphics[scale=0.7, angle=-0]{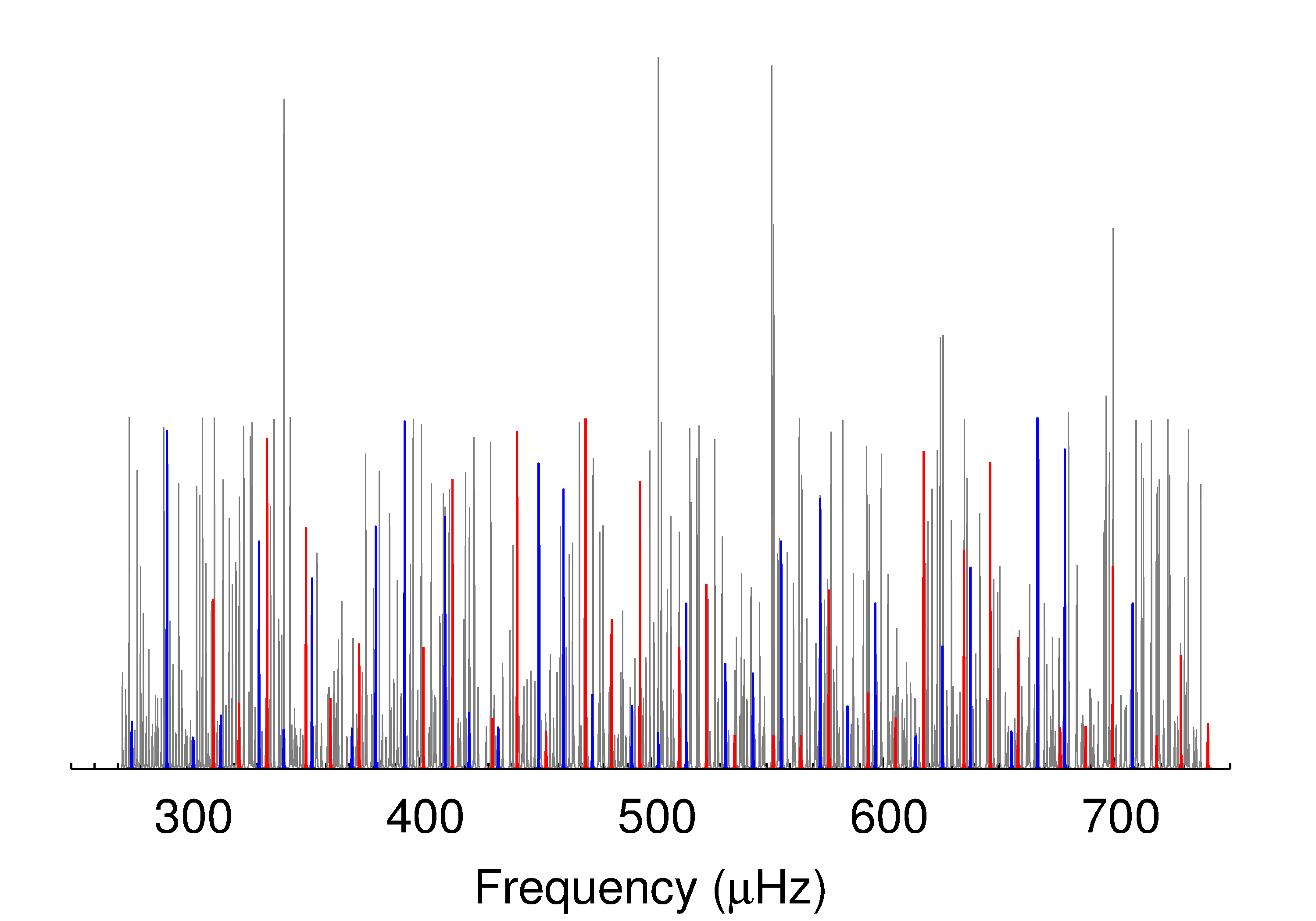}}
            \caption{\label{Fig_spectrum} Simulated spectrum of axisymmetric modes (only) for the model presented Sect. \ref{S_Ex}, it contains the eigenfrequencies of odd and even Island modes (in blue and red resp.), as well as modes trapped only in the $\mu$ gradient, Chaotic modes and Whisperring gallery modes in grey. }
\end{figure}

In this study, we have computed axisymmetric eigenmodes with frequencies ranging from 250 $\mu$Hz to 750 $\mu$Hz (10-30 $\Omega_k$), by scanning the frequency interval we made sure that all the axisymmetric modes with contributions up to $\ell=80$ are found. The resulting pulsation spectrum presents around 800 modes, which corresponds to one mode every 0.63 $\mu$Hz in average. It is shown in Fig. \ref{Fig_spectrum}, where we have simulated each frequency peak by a Lorentzian of width $0.05 \, \mu$Hz and height 1. The quantity on the y axis of that graph has no significance. The different {\it amplitudes} are to be attributed to the fact that when two frequencies are closer than $0.05 \, \mu$Hz, their {\it amplitudes} add up.

Fast rotation drastically modifies the geometry of the pulsation modes; their geometric structure determines their visibility, as well as the region they probe. Therefore, a correct characterisation of the geometry of different modes is an important issue for the asteroseismology of fast rotators. The geometry of each of the 800 modes have been visualised (based on their kinetic energy in a meridional plane) allowing a careful characterisation. Depending on the region of the frequency spectrum explored by our computations, four geometric classes are found.

At low frequency, we mostly find modes trapped in the region of sharp variation of the mean molecular weight, characterised by a sharp feature in the  Brunt-V\"ais\"al\"a frequency. These modes are counterparts of high order g-modes in the rapidly rotating case. As in the non-rotating case, their wave number is proportional to $\sqrt{\ell (\ell+1)}$. Therefore, since we include components with $\ell$ up to 80 here, g-modes with high $\ell$ components are very numerous, even in the high frequency range. Their amplitude is confined around the central region, they have negligible amplitude in outer layers, and we expect them to undergo radiative damping, therefore, they should not be detected observationally. 

At higher frequency, we find whispering gallery modes, chaotic modes and island modes such as those computed in homogeneous models of stars by \cite{Lignieres2006}. Whispering gallery modes are the counterpart of high degree acoustic modes in the rapidly rotating case. They probe the outer layers of the star, but as they show very low visibility factors, they might not be detected.

Chaotic modes have no counterparts in the non-rotating case. They have been computed only in very high rotating models of stars \citep{Lignieres2009}. Their spatial distribution does not show a simple symmetry. They have significant amplitude in the whole of the stellar interior. The lack of symmetry should result in a low cancellation factor. Therefore, these modes are expected to be detected observationally. 

Whispering gallery and chaotic modes correspond, respectively, to high and intermediate values of $\ell-|m|$ at zero rotation \citep{Lignieres2006}.

\begin{figure}[t!]
\centering 
  	    {\includegraphics[width=0.85\linewidth, angle=-0]{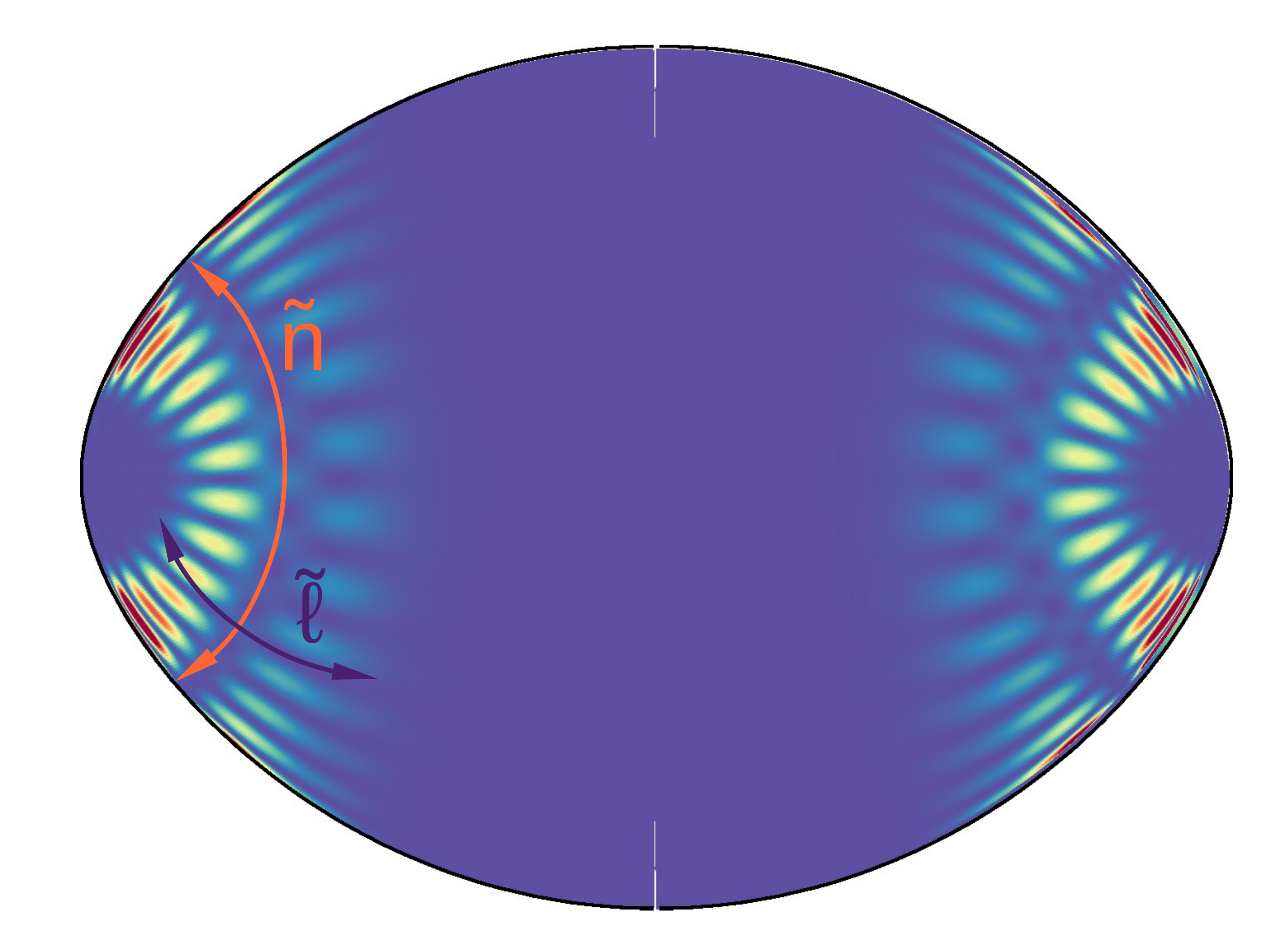}}
            \caption{\label{Fig_2D_Diag_ilot} Kinetic energy $\rho |v_{puls}|^2$ in a meridional plane, of an $\tilde{\ell}=1$ axisymmetric island mode. 
              }
\end{figure}

Island modes are counterparts of low degree acoustic modes in the rapidly rotating case. They probe outer layers of the star, and present good geometric visibility factors. Therefore they should be easily detected observationally. The geometry of such a mode is presented Fig.\ref{Fig_2D_Diag_ilot} with the distribution of kinetic energy of the mode in a meridional plane. As mentioned in \cite{Reese2008} if their non-rotating counterpart is characterised by the known quantum numbers $(n,\ell,m)$, the geometry of  these Island modes follows a different distribution characterised by a new set of quantum numbers $(\tilde{n} = 2 n + p, \tilde{\ell} = \frac{\ell+|m|+p}{2},m)$, where $p$ indicates the mode's parity. The two new quantum numbers $\tilde{\ell}$ and $\tilde{n}$ correspond respectively to the number of nodes along and perpendicular to the ray path (Fig.\ref{Fig_2D_Diag_ilot}), see \cite{Reese2008,Reese2009b}. 

\begin{figure}[t!]
\centering 
  	    {\includegraphics[width=0.78\linewidth, angle=-0]{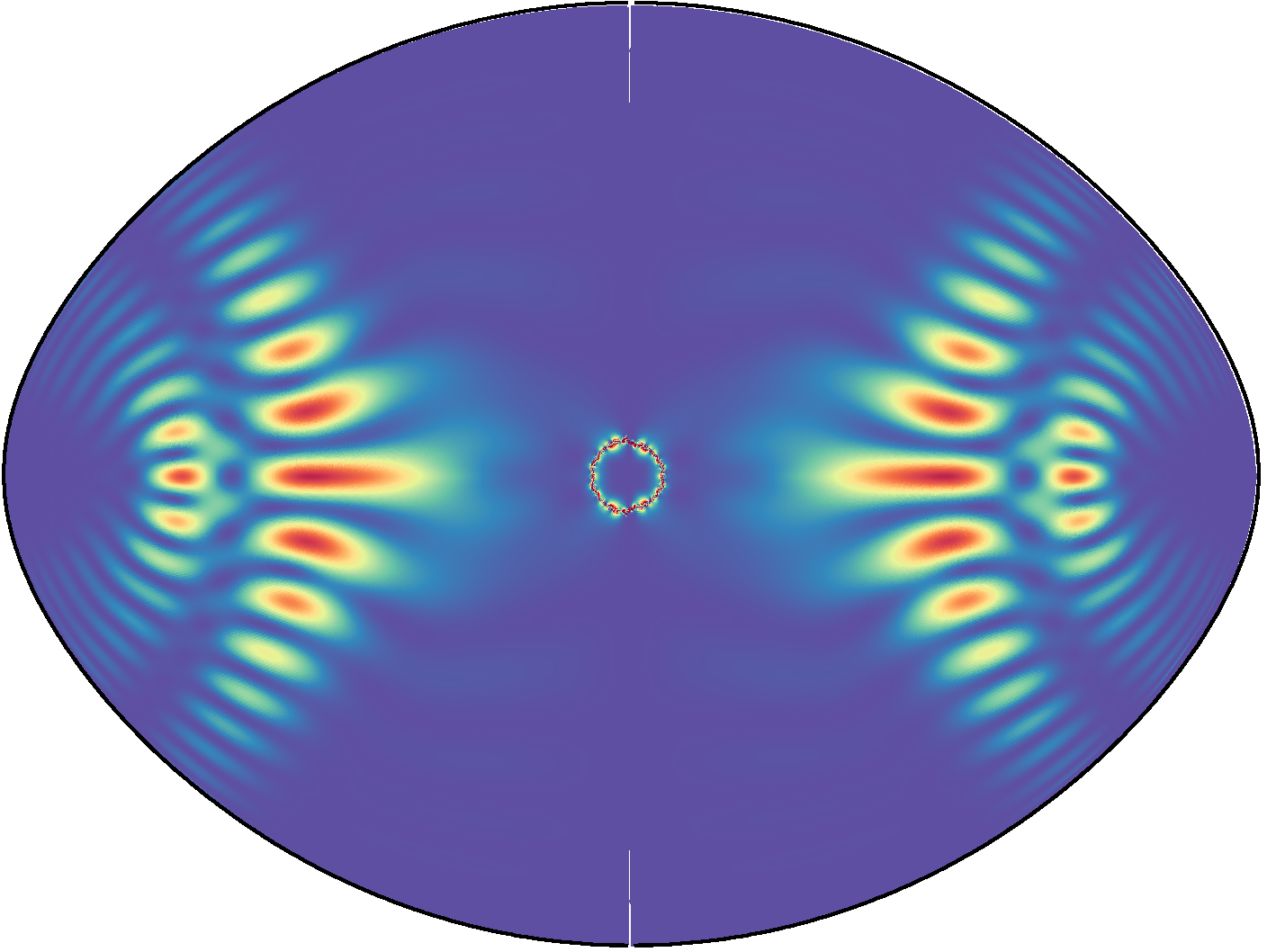}}\\
            \caption{\label{Fig_2D_Diag_mixte} Kinetic energy divided by the square root of the density and multiplied by the square of the distance to the axis in a meridional plane, for an axisymmetric mixed island mode of $\tilde{\ell}=1$.
              }
\end{figure}

Some island modes in the explored frequency range are found to be mixed, i.e.\ they behave as pressure modes in the envelope and gravity modes near the core. In Fig.\ref{Fig_2D_Diag_mixte} we give the geometry distribution of a mixed mode in a meridional plane, illustrated by the kinetic energy divided by the square root of the density and multiplied by the square of the distance to the axis. Mixed non-radial modes have been theoretically known since \cite{Osaki1975} discovered them in a non-rotating model of a 10 M$_{\odot}$ star on the main sequence. For the first time, we show that such gravito-acoustic mixed modes are present in the theoretical spectrum of rapidly rotating stellar models too. Indeed, as a star evolves, the Brunt-V\"ais\"al\"a frequency, i.e.\ the maximum frequency of trapped g-modes, increases owing to the increasing core density. Therefore, the frequencies of g modes become comparable to those of p modes, and some of them adopt a mixed character. These mixed modes are very useful because, unlike pure g modes, they have large enough surface amplitudes to be detected and yet are sensitive to the structure of the core. In the space photometry era, these mixed modes have been intensively observed in evolved solar type stars, first with CoRoT \citep[][]{Deheuvels2010} and with Kepler \citep[e.g.][]{Campante2011,Mathur2011}. These modes enabled one to probe the structure of the core of subgiants \citep{Deheuvels2011} and red giants \citep{Beck2011,Mosser2011}, thus making it possible to discriminate between evolutionary scenarios \citep[shell hydrogen or core helium burning, see for instance][]{Mosser2012a}. The detection of mixed modes that are split by rotation also allow to probe the rotation rate of stars even in their deepest interior \citep[][]{Beck2012,Deheuvels2012}. The situation of fast rotating intermediate mass stars is much less favourable, not only because they are not solar-like pulsators, but also because the impact of rapid rotation on their spectra was poorly modelled. Once stellar and pulsations models are adapted to such impact, the use of mixed modes as probe for rapidly rotating stars interior seems now reachable in the near future.  


\section{Island modes regularity}
\label{S_Reg}
\begin{figure}[t!]
\centering 
  	    {\includegraphics[scale=0.7, angle=-0]{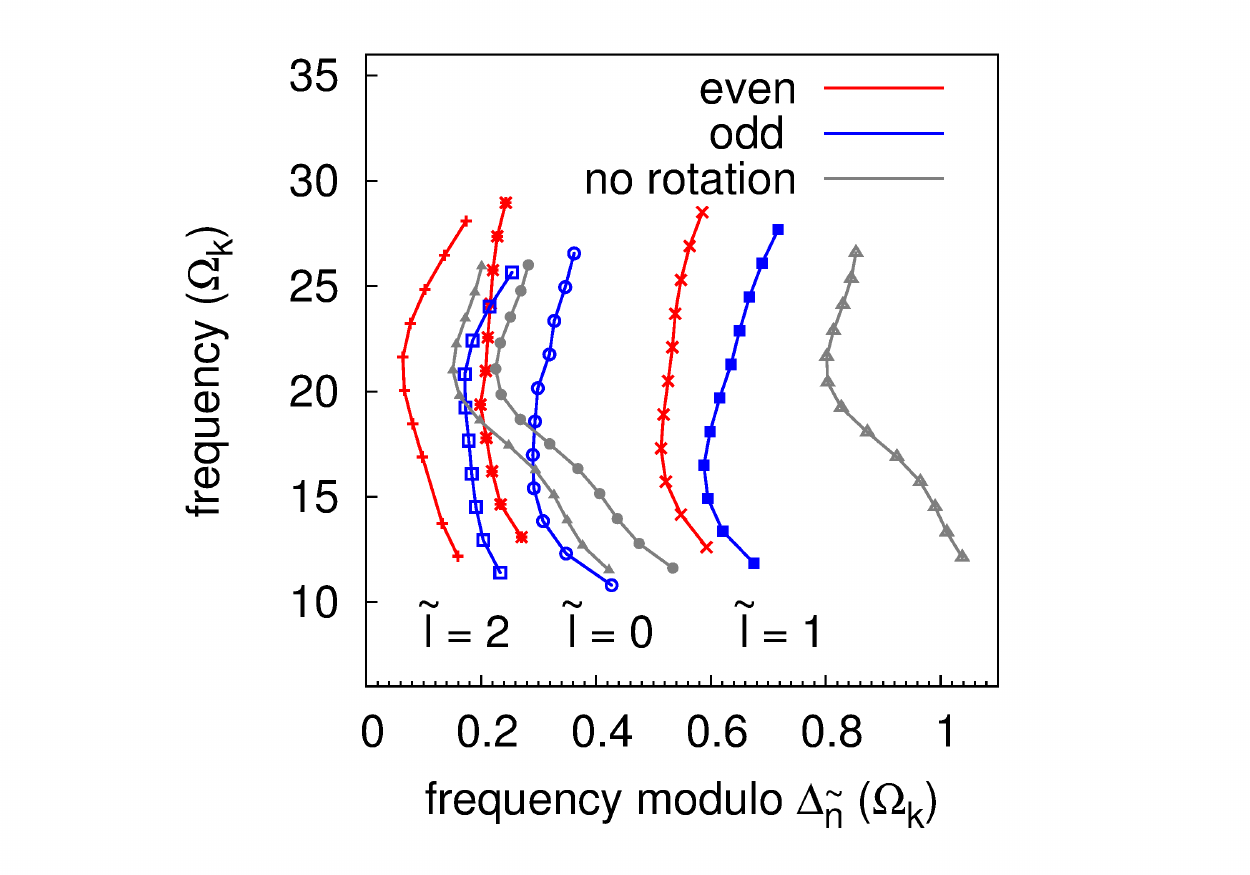}}
            \caption{\label{Fig_DiagEch} \'Echelle diagram for axisymmetric Island modes of different degree $\tilde{\ell}$: from left to right, $\tilde{\ell} = 2,0$ and 1, for odd (blue) and even modes (red).
              }
\end{figure}

Low $\tilde{\ell}$ island modes are expected to be the most visible modes in the seismic spectra of rapidly rotating stars  since they are the rotating counterparts to modes with low $\ell$ values. Their frequency organisation has been studied in \cite{Lignieres2006} in non-distorted polytropic models (of index n=3), \cite{Lignieres2008,Lignieres2009} using ray dynamics in polytropic models, \cite{Reese2008} in fully distorted polytropic models, \cite{Reese2009a} for homogeneous models distorted using the SCF method \citep{MacGregor2007}, and \cite{Pasek2012} who have developed an semi-analytic formula for their spectrum under the Cowling approximation, and neglecting the Coriolis force.  

The first difficulty with studying  the spectrum of island modes is to be able to identify  them among all the other type of modes present in the spectra of rapidly rotating stars (chaotic and whispering gallery modes). This was done by finding a few island modes in the high frequency regime, identifying their spatial geometry (i.e. $\tilde{n}, \, \tilde{\ell}$, m is fixed) determining their frequency spacings and then checking visually the modes which frequencies are around the expected one.  

\subsection{Asymptotic frequencies}
\cite{Reese2008} suggested a new asymptotic formula for island modes:
\begin{align}
\omega_{\tilde{n},\tilde{\ell},m}\, = \, \tilde{n} \,\Delta_{\tilde{n}} \,+ \,\tilde{\ell} \,\Delta_{\tilde{\ell}} \,+ \,m^2 \,\Delta_{m} \,+\, \tilde{\alpha}\,,
\label{freq_asympt}
\end{align}
which they have tested for the SCF homogeneous models. Based on our synthetic spectrum of axisymmetric modes we were able to determine the parameters $\Delta_{\tilde{n}}$, $\Delta_{\tilde{\ell}}$ and  $\tilde{\alpha}$ of this asymptotic formula based on a fit on our frequency spectrum. This fit has been carried out for the fully rotating 2d model, as well as for the spherical non-rotating model corresponding to the structural quantities of the 2d model at the latitude $\theta_m = \cos^{-1}(1/\sqrt{3})$. The results are shown in Table (\ref{tab_sep}).

\begin{table}[h!]
\caption{\label{tab_sep} Values of the parameters of Eq. (\ref{freq_asympt}) for the non-rotating model (first row), and for the 2D model (second row).}
\centering
\begin{tabular}{lccccc}
\hline
\hline
$\frac{\Omega}{\Omega_k}$  &  $\frac{\Delta_{\tilde{n}}}{\Omega_k}$  &  $\frac{\Delta_{\tilde{l}}}{\Omega_k}$  & $\frac{\tilde{\alpha}}{\Omega_k}$\\
\hline
0.000 & 1.184 & 0.523 & 1.058 \\
0.807 & 0.795 & 0.459 & 1.015\\
\hline
\end{tabular}
\end{table}

In order to highlight this regularity, one can also plot, what we call an {\it \'echelle diagram} which is the spectrum folded about the spacing $\Delta_{\tilde{n}}$. The result is shown in Fig. \ref{Fig_DiagEch}, which gives the \'echelle diagram for axisymmetric island modes of different parity and for $\tilde{\ell} = 0,1,$ and 2. For a given parity, the modes frequencies line up along ridges of given $\tilde{\ell}$ values. 

This feature is of course not surprising as island modes are acoustic modes, and once the shape of their cavity is known, their spectrum is determined by the stratification of the fluid in the cavity. 

As pointed out in non-rotating models of evolved low mass stars, when acoustic modes acquire a gravity contribution during an avoided crossing for instance, their frequencies have the tendency to deviate from the regular pattern defined by the large spacing. That is what is causing the deviation at low frequency in Fig. \ref{Fig_DiagEch}, where the corresponding eigenmodes clearly show trapping in the $\mu$ gradient region. For instance, the lowest frequency mode on the odd $\tilde{\ell} = 1$ ridge is the mixed mode whose kinetic energy distribution is given in Fig.\ref{Fig_2D_Diag_mixte}.

\subsection{Theoretical {\it Large Separation}}

\begin{figure}[t!]
\centering 
  	    {\includegraphics[width=0.78\linewidth, angle=-0]{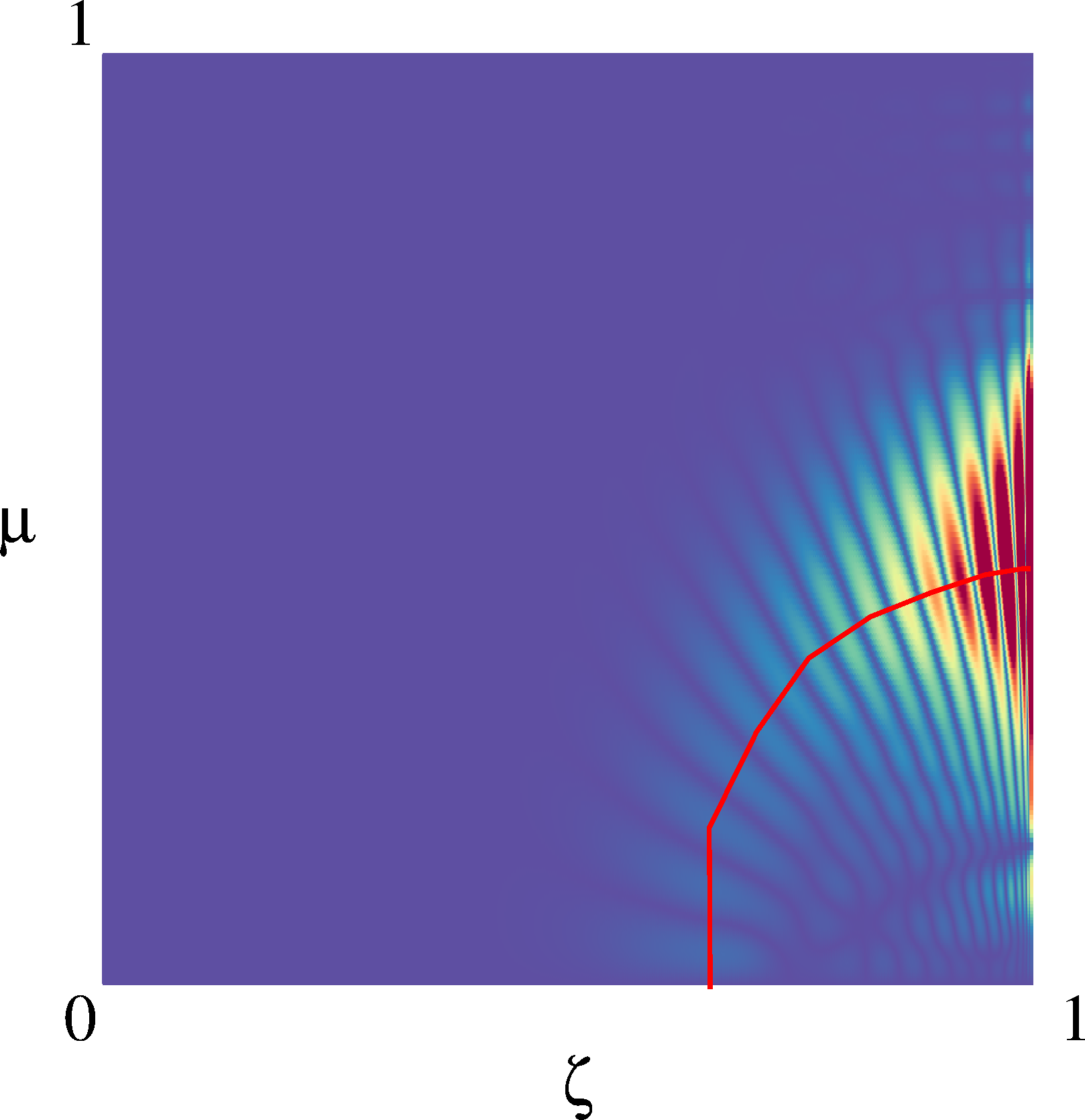}}\\
            \caption{\label{Fig_2D_charact} Kinetic energy distribution of an $\tilde{\ell} = 0$ axisymmetric island mode in the ($\zeta$, $\mu$) plan, where $\zeta$ is the pseudo-radial coordinate, and $\mu$ is related to the colatitude by: $\mu = \cos \theta$. The red curve connects the local maxima of the kinetic energy, that defines the ray path of island modes in the explored model.
              }
\end{figure}

In previous works of full modelling \citep{Reese2008,Reese2009b} the hypothesis that the pseudo-large separation $\Delta_{\tilde{n}}$ is a measure of the travel time in the island modes cavity has been mentioned but never actually tested. The difficulty here is to find the ray path for island modes geometry. The approach adopted consists in finding the underlying ray path for our model based on the topology of the 2-dimensional eigenfunctions. For the sake of simplicity, we assume that the underlying ray path corresponds to the local maxima of the pressure perturbation for an $\tilde{\ell} = 0$ island mode. This yields the $\mu(\zeta)$ curve plotted in red in Fig. \ref{Fig_2D_charact}, where $\mu=\cos\theta$ and $\zeta$ the pseudo-radial coordinate. The theoretical large separation is then given by:
\begin{equation}
\Delta_{\tilde{n}}^{theo}\, =\, \int_{ray} \frac{ds(\zeta)}{c_{s}}\,,
\end{equation}  
where $c_s$ is the sound speed along the ray, and $ds$ corresponds to the arclength along the ray. Such a method allowed us to determine a theoretical large frequency separation of $\Delta_{\tilde{n}}^{theo} = 0.829 \,\Omega_k$, which is compatible with the large separation computed from the theoretical seismic spectrum by averaging the individual values $\Delta_{\tilde{n}} = (0.795 \pm 0.012) \Omega_k$, but not within the dispersion. 

This is a rough estimate of the inverse of the travel time on the island modes ray paths, but this confirms that this new spacing $\Delta_{\tilde{n}}$ is an equivalent of the large separation $\Delta \nu$ of acoustic modes without rotation. Once island modes are identified in the spectrum, $\Delta_{\tilde{n}}$ allows one to constrain the sound speed profile in the outer radiative envelope of these rapidly rotating intermediate-mass stars. The other way around, in theoretical studies, it can also be of great help in order to track island modes in synthetic spectra by pointing toward the right pattern.

Most of the uncertainty of this method lies in the determination of the ray path from the eigenfunctions of the modes. Of course the most accurate way to compute the ray path is actually by studying the dynamics of acoustics rays \citep[as is done in ][for polytropes]{Lignieres2009}, but so far it has never been done in a realistic model, and without making certain assumptions such as that the Coriolis force doesn't affect too much acoustic-ray dynamics \citep[][have shown that its effects even on relatively low radial orders are weak]{Reese2006}, and the Cowling approximation. The presence of mixed modes in our spectrum of a realistic mid main sequence model requires one to drop these assumptions. We leave this study for future work.


\section{Conclusions}
\label{S_CCL}
A new modelling chain which computes non-radial adiabatic pulsations in rapidly rotating distorted models of stars has been presented. The first step consists in building a distorted stellar model starting with a spherically averaged one for arbitrary rotation laws \citep{Roxburgh2006}. The second step is the computation of its stellar pulsations by fully acounting for the impact of centrifugal distortion and for the effect of the Coriolis force on mode dynamics (using the 2D non-perturbative oscillation code ACOR, see Paper I). In rotationally distorted stars, the structural gradients are steeper near the poles than near the equator. To handle that, a new coordinate system was developed that greatly simplifies the computations, and disminishes numerical sources of uncertainties. That allowed to compute for the first time the synthetic spectrum of a realistic distorted model of a non-homogeneous star, evolved on its main sequence.

This approach has been applied to a 2 M$_{\odot}$ star evolved midway on its main sequence ($X_c = 0.35$). The simulations yield a wealth of modes with a spectrum denser than its non-rotating counter-part. Four types of modes have been computed, among which island modes that are expected to be the most visible ones. The regularity of island modes is confirmed in the general case of non-barotropic stratified stellar models. In particular the spectrum is organised around the large spacing $\Delta_{\tilde{n}}$, and it is possible to build \'echelle diagrams for island modes. Once the shape of the cavity is correctly defined, it is possible to retrieve a theoretical value of this large spacing as the inverse of the travel time along the modes path which is compatible with the value computed from the spectrum.

Moreover, we have shown that, just like their non-rotating counterparts, the island modes can be involved in avoided crossings with modes trapped near the core. Such mixed modes (in the sense of gravito-acoustic modes) deviate from the regular pattern defined by the large spacing $\Delta_{\tilde{n}}$.

Finally, several seismic diagnostics based on these island modes arise.
The large spacing of the island-mode spectrum is related to the stratification of the star in the outer radiative envelope.
The deviation from the regular spacing due to avoided crossings should allow to constrain the evolutionnary status of the star.

\begin{acknowledgements}
  RMO thanks D. Reese and F. Lignieres for fruitful discussions, as well as J. Christensen-Dalsgaard for careful reading which helped to improve the manuscript. Funding for the Stellar Astrophysics Centre is provided by The Danish National Research Foundation (Grant DNRF106). The research is supported by the ASTERISK project (ASTERoseismic Investigations with SONG and Kepler) funded by the European Research Council (Grant agreement no.: 267864). IWR gratefully acknowledges support from the Leverhulme Foundation under grant EM-2012-035/4.  
\end{acknowledgements}

\bibliographystyle{aa}

\bibliography{RMO_biblio}

\begin{thebibliography}{46}
\expandafter\ifx\csname natexlab\endcsname\relax\def\natexlab#1{#1}\fi

\bibitem[{{Alexander} \& {Ferguson}(1994)}]{Alexander1994}
{Alexander}, D.~R. \& {Ferguson}, J.~W. 1994, \apj, 437, 879

\bibitem[{{Angulo} {et~al.}(1999){Angulo}, {Arnould}, {Rayet}, {Descouvemont},
  {Baye}, {Leclercq-Willain}, {Coc}, {Barhoumi}, {Aguer}, {Rolfs}, {Kunz},
  {Hammer}, {Mayer}, {Paradellis}, {Kossionides}, {Chronidou}, {Spyrou},
  {degl'Innocenti}, {Fiorentini}, {Ricci}, {Zavatarelli}, {Providencia},
  {Wolters}, {Soares}, {Grama}, {Rahighi}, {Shotter}, \& {Lamehi
  Rachti}}]{Angulo1999}
{Angulo}, C., {Arnould}, M., {Rayet}, M., {et~al.} 1999, Nuclear Physics A,
  656, 3

\bibitem[{{Beck} {et~al.}(2011){Beck}, {Bedding}, {Mosser}, {Stello}, {Garcia},
  {Kallinger}, {Hekker}, {Elsworth}, {Frandsen}, {Carrier}, {De Ridder},
  {Aerts}, {White}, {Huber}, {Dupret}, {Montalb{\'a}n}, {Miglio}, {Noels},
  {Chaplin}, {Kjeldsen}, {Christensen-Dalsgaard}, {Gilliland}, {Brown},
  {Kawaler}, {Mathur}, \& {Jenkins}}]{Beck2011}
{Beck}, P.~G., {Bedding}, T.~R., {Mosser}, B., {et~al.} 2011, Science, 332, 205

\bibitem[{{Beck} {et~al.}(2012){Beck}, {Montalban}, {Kallinger}, {De Ridder},
  {Aerts}, {Garc{\'{\i}}a}, {Hekker}, {Dupret}, {Mosser}, {Eggenberger},
  {Stello}, {Elsworth}, {Frandsen}, {Carrier}, {Hillen}, {Gruberbauer},
  {Christensen-Dalsgaard}, {Miglio}, {Valentini}, {Bedding}, {Kjeldsen},
  {Girouard}, {Hall}, \& {Ibrahim}}]{Beck2012}
{Beck}, P.~G., {Montalban}, J., {Kallinger}, T., {et~al.} 2012, \nat, 481, 55

\bibitem[{{Campante} {et~al.}(2011){Campante}, {Handberg}, {Mathur},
  {Appourchaux}, {Bedding}, {Chaplin}, {Garc{\'{\i}}a}, {Mosser}, {Benomar},
  {Bonanno}, {Corsaro}, {Fletcher}, {Gaulme}, {Hekker}, {Karoff}, {R{\'e}gulo},
  {Salabert}, {Verner}, {White}, {Houdek}, {Brand{\~a}o}, {Creevey}, {Do{\v
  g}an}, {Bazot}, {Christensen-Dalsgaard}, {Cunha}, {Elsworth}, {Huber},
  {Kjeldsen}, {Lundkvist}, {Molenda-{\.Z}akowicz}, {Monteiro}, {Stello},
  {Clarke}, {Girouard}, \& {Hall}}]{Campante2011}
{Campante}, T.~L., {Handberg}, R., {Mathur}, S., {et~al.} 2011, \aap, 534, A6

\bibitem[{{Clement}(1978)}]{Clement1978}
{Clement}, M.~J. 1978, \apj, 222, 967

\bibitem[{{Deheuvels} {et~al.}(2010){Deheuvels}, {Bruntt}, {Michel}, {Barban},
  {Verner}, {R{\'e}gulo}, {Mosser}, {Mathur}, {Gaulme}, {Garcia}, {Boumier},
  {Appourchaux}, {Samadi}, {Catala}, {Baudin}, {Baglin}, {Auvergne},
  {Roxburgh}, \& {P{\'e}rez Hern{\'a}ndez}}]{Deheuvels2010}
{Deheuvels}, S., {Bruntt}, H., {Michel}, E., {et~al.} 2010, \aap, 515, A87+

\bibitem[{{Deheuvels} {et~al.}(2012){Deheuvels}, {Garc{\'{\i}}a}, {Chaplin},
  {Basu}, {Antia}, {Appourchaux}, {Benomar}, {Davies}, {Elsworth}, {Gizon},
  {Goupil}, {Reese}, {Regulo}, {Schou}, {Stahn}, {Casagrande},
  {Christensen-Dalsgaard}, {Fischer}, {Hekker}, {Kjeldsen}, {Mathur}, {Mosser},
  {Pinsonneault}, {Valenti}, {Christiansen}, {Kinemuchi}, \&
  {Mullally}}]{Deheuvels2012}
{Deheuvels}, S., {Garc{\'{\i}}a}, R.~A., {Chaplin}, W.~J., {et~al.} 2012, \apj,
  756, 19

\bibitem[{{Deheuvels} \& {Michel}(2011)}]{Deheuvels2011}
{Deheuvels}, S. \& {Michel}, E. 2011, \aap, 535, A91

\bibitem[{{Dziembowski} \& {Goode}(1992)}]{Dziembowski1992}
{Dziembowski}, W.~A. \& {Goode}, P.~R. 1992, \apj, 394, 670

\bibitem[{{Eddington}(1925)}]{Eddington1925}
{Eddington}, A.~S. 1925, The Observatory, 48, 73

\bibitem[{{Espinosa Lara} \& {Rieutord}(2013)}]{EspinosaLara2013}
{Espinosa Lara}, F. \& {Rieutord}, M. 2013, \aap, 552, A35

\bibitem[{{Hurley} \& {Roberts}(1964)}]{Hurley1964}
{Hurley}, M. \& {Roberts}, P.~H. 1964, \apj, 140, 583

\bibitem[{{Iglesias} \& {Rogers}(1996)}]{Iglesias1996}
{Iglesias}, C.~A. \& {Rogers}, F.~J. 1996, \apj, 464, 943

\bibitem[{{Jackson}(1970)}]{Jackson1970}
{Jackson}, S. 1970, \apj, 161, 579

\bibitem[{{Jackson} {et~al.}(2004){Jackson}, {MacGregor}, \&
  {Skumanich}}]{Jackson2004}
{Jackson}, S., {MacGregor}, K.~B., \& {Skumanich}, A. 2004, \apj, 606, 1196

\bibitem[{{James}(1964)}]{James1964}
{James}, R.~A. 1964, \apj, 140, 552

\bibitem[{{Ledoux}(1951)}]{Ledoux1951}
{Ledoux}, P. 1951, \apj, 114, 373

\bibitem[{{Ligni{\`e}res} \& {Georgeot}(2008)}]{Lignieres2008}
{Ligni{\`e}res}, F. \& {Georgeot}, B. 2008, \pre, 78, 016215

\bibitem[{{Ligni{\`e}res} \& {Georgeot}(2009)}]{Lignieres2009}
---. 2009, \aap, 500, 1173

\bibitem[{{Ligni{\`e}res} {et~al.}(2006){Ligni{\`e}res}, {Rieutord}, \&
  {Reese}}]{Lignieres2006}
{Ligni{\`e}res}, F., {Rieutord}, M., \& {Reese}, D. 2006, \aap, 455, 607

\bibitem[{{MacGregor} {et~al.}(2007){MacGregor}, {Jackson}, {Skumanich}, \&
  {Metcalfe}}]{MacGregor2007}
{MacGregor}, K.~B., {Jackson}, S., {Skumanich}, A., \& {Metcalfe}, T.~S. 2007,
  \apj, 663, 560

\bibitem[{{Mathis} \& {Zahn}(2004)}]{Mathis2004}
{Mathis}, S. \& {Zahn}, J. 2004, \aap, 425, 229

\bibitem[{{Mathur} {et~al.}(2011){Mathur}, {Handberg}, {Campante},
  {Garc{\'{\i}}a}, {Appourchaux}, {Bedding}, {Mosser}, {Chaplin}, {Ballot},
  {Benomar}, {Bonanno}, {Corsaro}, {Gaulme}, {Hekker}, {R{\'e}gulo},
  {Salabert}, {Verner}, {White}, {Brand{\~a}o}, {Creevey}, {Do{\v g}an},
  {Elsworth}, {Huber}, {Hale}, {Houdek}, {Karoff}, {Metcalfe},
  {Molenda-{\.Z}akowicz}, {Monteiro}, {Thompson}, {Christensen-Dalsgaard},
  {Gilliland}, {Kawaler}, {Kjeldsen}, {Quintana}, {Sanderfer}, \&
  {Seader}}]{Mathur2011}
{Mathur}, S., {Handberg}, R., {Campante}, T.~L., {et~al.} 2011, \apj, 733, 95

\bibitem[{{Monaghan} \& {Roxburgh}(1965)}]{Monaghan1965}
{Monaghan}, J.~J. \& {Roxburgh}, I.~W. 1965, \mnras, 131, 13

\bibitem[{{Mosser} {et~al.}(2011){Mosser}, {Barban}, {Montalb{\'a}n}, {Beck},
  {Miglio}, {Belkacem}, {Goupil}, {Hekker}, {De Ridder}, {Dupret}, {Elsworth},
  {Noels}, {Baudin}, {Michel}, {Samadi}, {Auvergne}, {Baglin}, \&
  {Catala}}]{Mosser2011}
{Mosser}, B., {Barban}, C., {Montalb{\'a}n}, J., {et~al.} 2011, \aap, 532, A86

\bibitem[{{Mosser} {et~al.}(2012){Mosser}, {Goupil}, {Belkacem}, {Michel},
  {Stello}, {Marques}, {Elsworth}, {Barban}, {Beck}, {Bedding}, {De Ridder},
  {Garc{\'{\i}}a}, {Hekker}, {Kallinger}, {Samadi}, {Stumpe}, {Barclay}, \&
  {Burke}}]{Mosser2012a}
{Mosser}, B., {Goupil}, M.~J., {Belkacem}, K., {et~al.} 2012, \aap, 540, A143

\bibitem[{{Osaki}(1975)}]{Osaki1975}
{Osaki}, J. 1975, \pasj, 27, 237

\bibitem[{{Ostriker} \& {Mark}(1968)}]{Ostriker1968}
{Ostriker}, J.~P. \& {Mark}, J. 1968, \apj, 151, 1075

\bibitem[{{Ouazzani} {et~al.}(2012){Ouazzani}, {Dupret}, \&
  {Reese}}]{Ouazzani2012b}
{Ouazzani}, R.-M., {Dupret}, M.-A., \& {Reese}, D.~R. 2012, \aap, 547, A75

\bibitem[{{Pasek} {et~al.}(2012){Pasek}, {Ligni{\`e}res}, {Georgeot}, \&
  {Reese}}]{Pasek2012}
{Pasek}, M., {Ligni{\`e}res}, F., {Georgeot}, B., \& {Reese}, D.~R. 2012, \aap,
  546, A11

\bibitem[{{Reese} {et~al.}(2006){Reese}, {Ligni{\`e}res}, \&
  {Rieutord}}]{Reese2006}
{Reese}, D., {Ligni{\`e}res}, F., \& {Rieutord}, M. 2006, \aap, 455, 621

\bibitem[{{Reese} {et~al.}(2008){Reese}, {Ligni{\`e}res}, \&
  {Rieutord}}]{Reese2008}
---. 2008, \aap, 481, 449

\bibitem[{{Reese} {et~al.}(2009{\natexlab{a}}){Reese}, {MacGregor}, {Jackson},
  {Skumanich}, \& {Metcalfe}}]{Reese2009a}
{Reese}, D.~R., {MacGregor}, K.~B., {Jackson}, S., {Skumanich}, A., \&
  {Metcalfe}, T.~S. 2009{\natexlab{a}}, \aap, 506, 189

\bibitem[{{Reese} {et~al.}(2009{\natexlab{b}}){Reese}, {Thompson}, {MacGregor},
  {Jackson}, {Skumanich}, \& {Metcalfe}}]{Reese2009b}
{Reese}, D.~R., {Thompson}, M.~J., {MacGregor}, K.~B., {et~al.}
  2009{\natexlab{b}}, \aap, 506, 183

\bibitem[{{Rieutord}(2007)}]{Rieutord2007}
{Rieutord}, M. 2007, ArXiv Astrophysics e-prints: arXiv:astro-ph/0702384

\bibitem[{{Rogers} \& {Nayfonov}(2002)}]{Rogers2002}
{Rogers}, F.~J. \& {Nayfonov}, A. 2002, \apj, 576, 1064

\bibitem[{{Roxburgh}(2004)}]{Roxburgh2004}
{Roxburgh}, I.~W. 2004, \aap, 428, 171

\bibitem[{{Roxburgh}(2006)}]{Roxburgh2006}
---. 2006, \aap, 454, 883

\bibitem[{{Roxburgh}(2008)}]{Roxburgh2008}
---. 2008, \apss, 316, 75

\bibitem[{{Roxburgh} {et~al.}(1965){Roxburgh}, {Griffith}, \&
  {Sweet}}]{Roxburgh1965}
{Roxburgh}, I.~W., {Griffith}, J.~S., \& {Sweet}, P.~A. 1965, \zap, 61, 203

\bibitem[{{Scuflaire} {et~al.}(2008){Scuflaire}, {Montalb{\'a}n}, {Th{\'e}ado},
  {Bourge}, {Miglio}, {Godart}, {Thoul}, \& {Noels}}]{Scuflaire2008}
{Scuflaire}, R., {Montalb{\'a}n}, J., {Th{\'e}ado}, S., {et~al.} 2008, \apss,
  316, 149

\bibitem[{{Soufi} {et~al.}(1998){Soufi}, {Goupil}, \&
  {Dziembowski}}]{Soufi1998}
{Soufi}, F., {Goupil}, M.~J., \& {Dziembowski}, W.~A. 1998, \aap, 334, 911

\bibitem[{{Su{\'a}rez} {et~al.}(2010){Su{\'a}rez}, {Goupil}, {Reese}, {Samadi},
  {Ligni{\`e}res}, {Rieutord}, \& {Lochard}}]{Suarez2010}
{Su{\'a}rez}, J.~C., {Goupil}, M.~J., {Reese}, D.~R., {et~al.} 2010, \apj, 721,
  537

\bibitem[{{Vorontsov}(1981)}]{Vorontsov1981}
{Vorontsov}, S.~V. 1981, \sovast, 25, 724

\bibitem[{{Vorontsov}(1983)}]{Vorontsov1983}
---. 1983, \solphys, 82, 379

\end{thebibliography}

\end{document}